\documentclass[12pt]{article}
\usepackage{amsfonts}
\usepackage{amsmath}

\advance \topmargin by -\headheight
\advance \topmargin by -\headsep
     
\evensidemargin \oddsidemargin
 \date{}    
\begin{document}

\topmargin -.6in
\def\rf#1{(\ref{eq:#1})}
\def\lab#1{\label{eq:#1}}
\def\nonu{\nonumber}
\def\br{\begin{eqnarray}}
\def\er{\end{eqnarray}}
\def\be{\begin{equation}}
\def\ee{\end{equation}}
\def\eq{\!\!\!\! &=& \!\!\!\! }
\def\ba{\be\begin{array}{c}}
\def\ea{\end{array}\ee}
\def\foot#1{\footnotemark\footnotetext{#1}}
\def\lb{\lbrack}
\def\rb{\rbrack}
\def\llangle{\left\langle}
\def\rrangle{\right\rangle}
\def\blangle{\Bigl\langle}
\def\brangle{\Bigr\rangle}
\def\llb{\left\lbrack}
\def\rrb{\right\rbrack}
\def\Blb{\Bigl\lbrack}
\def\Brb{\Bigr\rbrack}
\def\lcurl{\left\{}
\def\rcurl{\right\}}
\def\({\left(}
\def\){\right)}
\def\v{\vert}                     
\def\bv{\bigm\vert}               
\def\lskip{\vskip\baselineskip\vskip-\parskip\noindent}
\def\mskp{\par\vskip 0.3cm \par\noindent}
\def\sskp{\par\vskip 0.15cm \par\noindent}
\def\bc{\begin{center}}
\def\ec{\end{center}}

\def\tr{\mathop{\rm tr}}                  
\def\Tr{\mathop{\rm Tr}}                  
\makeatletter
\newcommand{\rd}{\@ifnextchar^{\DIfF}{\DIfF^{}}}
\def\DIfF^#1{%
   \mathop{\mathrm{\mathstrut d}}%
   \nolimits^{#1}\gobblespace}
\def\gobblespace{\futurelet\diffarg\opspace}
\def\opspace{%
   \let\DiffSpace\!%
   \ifx\diffarg(%
   \let\DiffSpace\relax
   \else
   \ifx\diffarg[%
   \let\DiffSpace\relax
   \else
   \ifx\diffarg\{%
   \let\DiffSpace\relax
   \fi\fi\fi\DiffSpace}
\newcommand{\deriv}[3][]{\frac{\rd^{#1}#2}{\rd #3^{#1}}}
\providecommand*{\dder}[3][]{%
\frac{\rd^{#1}#2}{\rd #3^{#1}}}
\providecommand*{\pder}[3][]{%
\frac{\partial^{#1}#2}{\partial #3^{#1}}}
\newcommand{\renewoperator}[3]{\renewcommand*{#1}{\mathop{#2}#3}}
\renewoperator{\Re}{\mathrm{Re}}{\nolimits}
\renewoperator{\Im}{\mathrm{Im}}{\nolimits}
\providecommand*{\iu}%
{\ensuremath{\mathrm{i}\,}}
\providecommand*{\eu}%
{\ensuremath{\mathrm{e}}}
\def\a{\alpha}
\def\b{\beta}
\def\c{\chi}
\def\d{\delta}
\def\D{\Delta}
\def\eps{\epsilon}
\def\vareps{\varepsilon}
\def\g{\gamma}
\def\G{\Gamma}
\def\grad{\nabla}
\newcommand{\h}{\frac{1}{2}}
\def\l{\lambda}
\def\om{\omega}
\def\s{\sigma}
\def\O{\Omega}
\def\p{\phi}
\def\vp{\varphi}
\def\P{\Phi}
\def\pa{\partial}
\def\pr{\prime}
\def\ti{\tilde}
\def\wti{\widetilde}
\newcommand{\cA}{\mathcal{A}}
\newcommand{\cB}{\mathcal{B}}
\newcommand{\cC}{\mathcal{C}}
\newcommand{\cD}{\mathcal{D}}
\newcommand{\cE}{\mathcal{E}}
\newcommand{\cF}{\mathcal{F}}
\newcommand{\cG}{\mathcal{G}}
\newcommand{\cH}{\mathcal{H}}
\newcommand{\cI}{\mathcal{I}}
\newcommand{\cJ}{\mathcal{J}}
\newcommand{\cK}{\mathcal{K}}
\newcommand{\cL}{\mathcal{L}}
\newcommand{\cM}{\mathcal{M}}
\newcommand{\cN}{\mathcal{N}}
\newcommand{\cO}{\mathcal{O}}
\newcommand{\cP}{\mathcal{P}}
\newcommand{\cQ}{\mathcal{Q}}
\newcommand{\cR}{\mathcal{R}}
\newcommand{\cS}{\mathcal{S}}
\newcommand{\cT}{\mathcal{T}}
\newcommand{\cU}{\mathcal{U}}
\newcommand{\cV}{\mathcal{V}}
\newcommand{\cW}{\mathcal{W}}
\newcommand{\cX}{\mathcal{X}}
\newcommand{\cY}{\mathcal{Y}}
\newcommand{\cZ}{\mathcal{Z}}
\newcommand{\nit}{\noindent}
\newcommand{\ct}[1]{\cite{#1}}
\newcommand{\bi}[1]{\bibitem{#1}}

\begin{center}
{\large\bf  On negative flows of the AKNS hierarchy and a class}
\end{center}
\begin{center}
{\large\bf  of deformations of bihamiltonian structure of hydrodynamic
type   }
\end{center}
\normalsize
\vskip .4in

\begin{center}
 H. Aratyn

\par \vskip .1in \noindent
Department of Physics \\
University of Illinois at Chicago\\
845 W. Taylor St.\\
Chicago, Illinois 60607-7059\\
\par \vskip .3in

\end{center}

\begin{center}
J.F. Gomes and A.H. Zimerman

\par \vskip .1in \noindent
Instituto de F\'{\i}sica Te\'{o}rica-UNESP\\
Rua Pamplona 145\\
01405-900 S\~{a}o Paulo, Brazil
\par \vskip .3in

\end{center}

\begin{abstract}
A deformation parameter of a bihamiltonian structure of hydrodynamic type
is shown to parameterize different extensions of the AKNS hierarchy to 
include negative flows.
This construction establishes a purely algebraic link between,
on the one hand, two realizations of the first negative flow of the AKNS model 
and, on the other, two-component generalizations
of Camassa-Holm and Dym type equations.

The two-component generalizations of Camassa-Holm and Dym type equations 
can be obtained from the negative order Hamiltonians constructed 
from the Lenard relations recursively applied
on the Casimir of the first Poisson bracket of hydrodynamic type.
The positive order Hamiltonians, which follow from Lenard scheme
applied on the Casimir of the second Poisson bracket of hydrodynamic type,
are shown to coincide with the Hamiltonians of the AKNS model.
The AKNS Hamiltonians give rise to charges conserved with respect to 
equations of motion of two-component Camassa-Holm 
and two-component Dym type equations.

\end{abstract}

\section{\sf Introduction}
Recently,  the celebrated shallow-water equation obtained by Camassa and Holm \ct{CH} 
\be
u_t -u_{txx}= - 3u u_x + 2 u_x u_{xx}  + u u_{xxx} -\kappa u_x,\quad
\kappa={\rm const}
\lab{ch1}
\ee
was extended in \ct{CH2} by adding on the right hand side a term $\rho \rho_x$
with a new variable $\rho$, which satisfies the continuity equation
$\rho_t+ (u\rho)_x=0$. 
The model resulting from the above generalization first
appeared  in the study of 
deformations of the bihamiltonian structure of hydrodynamic type 
\ct{LiuZhang,falqui} and was coined 2-component Camassa-Holm equation. 
Soon after its derivation the model was identified with
the first negative flow of the AKNS hierarchy \ct{CH2}.

Another well-known integrable partial differential equation of interest to our study 
is the Dym-type equation \ct{kruskal,cewen,HZh,alber1,alber2,
Brunelli:2003du}:
\be
-u_{xxt}= 2u_x u_{xx}+ u u_{xxx}- \kappa u_x  \, .
\lab{dym}
\ee
It can also be extended to two-component
version by adding a term $\rho \rho_x$ on the right hand side
of \rf{dym}. The resulting two variable system is shown here to be 
equivalent to the negative flow of one of the extensions of the AKNS model.
It is also equivalent to the special limiting procedure
of deformations of the bihamiltonian structure of hydrodynamic type.

In this paper the following is accomplished:
First, we explore the Schr\"{o}dinger spectral problem 
of second order describing both 2-component 
Camassa-Holm and Dym type equation for different values of the deformation
parameter $\mu$.
We show that  this Schr\"{o}dinger spectral problem can be cast into the 
linear  $2 \times 2$ matrix spectral problem. 
Using sl$(2)$ gauge invariance we transform the time evolution flow
of the linear spectral problem into the AKNS first negative flow.
We should point out that there exist several ways to extend the 
AKNS hierarchy to incorporate negative flows. These extensions
are parametrized by a single parameter identified with $\mu$. 
We associate two different constructions of the negative flows of the
AKNS hierarchy to the 2-component Camassa-Holm (for $\mu=1$)
and Dym type equation (for $\mu=0$).
The result concerning the 2-component Camassa-Holm
equation constitutes an algebraic version of the proof given in \ct{CH2}.
Using connections of the AKNS and deformed Sinh-Gordon models to
the 2-component Camassa-Holm and Dym type equations, respectively,
we are able to find explicit soliton solutions given in hodographic variables.
The relation to the AKNS models allows us to 
construct a new chain of charges conserved with respect to 
equations of motion of two-component Camassa-Holm 
and two-component Dym type equations.
For both hierarchies 
the modified AKNS Hamiltonians 
provide a tower of positive order Hamiltonians
obtained via the underlying Lenard relations
of the Poisson brackets of hydrodynamic type from the Casimir of the second
bracket.

In section \ref{section:akns}, we briefly review the algebraic approach to the 
AKNS model and show how to extend the model in two different ways to negative time flows based on the
zero-curvature identities. 
In section \ref{section:schroedinger}, we set up a class of two-component
Schr\"{o}dinger spectral problems parametrized by $\mu$.
In the next section \ref{section:gauge}, we transform the Schr\"{o}dinger 
spectral problem by the reciprocal transformation and 
linearize it.
The resulting linear $2 \times 2$ matrix spectral problem is
then transformed by an sl$(2)$ gauge transformation into the AKNS
Lax spectral problem. The time flows of 2-component Camassa-Holm and Dym
type equation are shown to coincide with two different negative flows of the extended
AKNS model. 
Our construction allows us to find, in subsection \ref{subsection:examples},
explicit soliton solutions for various values of $\mu$.
In section \ref{section:bihamiltonian}, we reproduce equations
of motion for $\mu\ne0$ and $\mu=0$ cases in the setting of deformations
of the bihamiltonian structure of hydrodynamic type.
Remarkably, the Hamiltonians governing positive evolution flows of the AKNS 
hierarchy define conserved charges for the 2-component Camassa-Holm and Dym
type equations.
Also the conserved charges induced by the AKNS model
satisfy among themselves the Lenard relations of the bihamiltonian structure of
hydrodynamic type.
Thus, the Hamiltonians of the bihamiltonian structure of
hydrodynamic type connected to 2-component Camassa-Holm and Dym
type equations split into two chains,
one of the positive order induced by the AKNS hierarchy and one of the 
negative order containing generators of the equations of motion
defining both hierarchies.

\section{\sf Extended AKNS model}
\label{section:akns}
First, let us present the AKNS hierarchy in the setting 
of the sl$(2)$ loop algebra endowed with homogeneous gradation 
defined by the operator $\l \rd / \rd \l$. A variable $\l$ plays 
a double role
of a loop parameter of the loop algebra and a spectral parameter
of the underlying hierarchy.
The matrix Lax operator $L$ for the AKNS hierarchy reads:
\be
L = \pder{}{y} - \begin{bmatrix} \l & 0 \\ 0& - \l \end{bmatrix} 
- \begin{bmatrix} 0 & q \\ r& 0 \end{bmatrix}, 
\lab{lopsl2}
\ee
where $\pa /\pa y$ is the derivative with respect to ``space'' variable $y$.
The matrix Lax operator can be compactly written as
$L= \pa /\pa y - E - A_0$, with $E= \l \sigma_{3}$ 
and the matrix $A_0=q \sigma_{+}+r \sigma_{-}$, 
where $\s_3$ is the Pauli matrix  and $\s_{\pm}$ are given in terms 
of other Pauli matrices $\s_1,\s_2$
\[
\s_{-}= \h \(\s_1 - \iu \s_2\)= \begin{bmatrix}
0 & 0 \\ 1 & 0 \end{bmatrix}, \quad
\s_{+}= \h \(\s_1 + \iu \s_2\)= \begin{bmatrix}
0 & 1 \\ 0 & 0 \end{bmatrix}\,.
\]
We work within an algebraic approach to the integrable models based on
the linear spectral problem $L (\Psi)=0$, which simplifies
considerably under a dressing  
transformation:
\be
\Theta^{-1 } \(\pder{}{ y} - E - A_0\)\Theta=\pder{}{y} - E  \, ,
\lab{dreslaxsl2}
\ee
where the dressing matrix 
$\Theta = \exp \( \sum_{i<0} \l^i \theta^{(i)} \) $
is an exponential in negative powers of the spectral parameter $\l$
on a formal loop space of sl$(2)$.
Similarly, for higher flows we obtain
\be
\Theta^{-1} \( \pder{}{t_n}- E^{(n)}- \sum_{i=0}^{n-1} \l^i D^{(i)}_n \) \Theta
= \pder{}{t_n}- E^{(n)}\, ,
\lab{dresbn}
\ee
where $E^{(n)}= \l^n \s_3$ and terms $D^{(i)}_n$ are obtained from 
projection $(\Theta E^{(n)} \Theta^{-1})_+$ of $\Theta E^{(n)} \Theta^{-1}$ 
on the positive powers of $\l$ via expansion relation:
\[ \(\Theta E^{(n)} \Theta^{-1}\)_{+} =
E^{(n)}+ \sum_{i=0}^{n-1} \l^i D^{(i)}_n \,.
\]
These dressing relations give rise to the zero-curvature conditions
for the positive flows of the AKNS hierarchy
\be
\left[ \pder{}{ y}- E - A_0, \pder{}{t_n}- E^{(n)}- \sum_{i=0}^{n-1} D_n^{(i)}
\right]
= \Theta \left[ \pder{}{y}- E\, ,\, \pder{}{t_n}- E^{(n)} \right]
\Theta^{-1 }  =0 \, .
\lab{z-curva}
\ee
In particular, for $n=2$ we obtain the second flow of the AKNS hierarchy:
\be
\pder{ r}{t_2} = -\h r_{yy}+ q\,r^2 \;\;\; ; \;\;\; \pder{q}{t_2} = \h
q_{yy}- q^2\,r \, ,
\lab{secflow}
\ee
which reproduces the familiar vector non-linear Schr\"{o}dinger equation.

According to \ct{kluwer},
the Hamiltonian densities of the AKNS model are defined as 
\be
{\cal H}_n= -\tr \(E^{(0)} A^{(-n)} \) = \frac{1}{2} \sum_{k=0}^{n-1} \tr \(
A^{(-k)}A^{(1+k-n)}\)\, ,
\lab{17a}
\ee
where $A^{(-n)}$ are given by
\[ 
\Theta_y \, \Theta^{-1} = \sum_{k=1}^{\infty} A^{(-k)} \l^{-k} \,,
\]
where the symbol $\tr$ in expression \rf{17a} denotes a sl$(2)$ 
trace.
We list the first two Hamiltonians. 
Inserting  $n=1$ in \rf{17a} we obtain:
\be
{\cal H}_1 = -\tr (E^{(0)} A^{(-1)} ) = \frac{1}{2}\tr ( A_0^2)
=r q \, .
\lab{18a}
\ee
Similarly, for $n=2$ we obtain 
\be
{\cal H}_2=  q r_y \, .
\lab{20a}
\ee
Next, we extend the AKNS model to include negative grade time evolution 
equations governed by the zero-curvature equations \ct{Aratyn:2000wr}
\be
\left[ \pder{}{ y}- E - A_0\, , \,\pder{}{t_{-n}}- D^{(-1)}-D^{(-2)}-\cdots  -
D^{(-n)}
\right]=0 \, .
\lab{t-j}
\ee
Here, we only consider the first negative flow with $n=1$ and
set for brevity $s=t_{-1}$. In this case the compatibility equation 
\rf{t-j} reduces to 
\br
(A_0)_s - D^{(-1)}_y + \left[ E+A_0 , D^{(-1)} \right]=0 \, .
\lab{t-1}
\er
A general solution of the compatibility equation \rf{t-1} is given by
\be
 D^{(-1)} = B \cE^{(-1)} B^{-1}, \quad A_0 = B_y \, B^{-1}\, ,
 \lab{lrakns}
\ee
in terms of the zero-grade group element, $B$, of ${\rm SL}(2)$, that satisfies 
equation: 
\be 
(B_y \, B^{-1})_s =
\left[  B \cE^{(-1)} B^{-1},E \right] 
\lab{ls}
\ee
or, equivalently,
\be
(B^{-1} B_s)_y
=\left[ \cE^{(-1)}, B^{-1}E B \right] \, .
\lab{ls-a}
\ee
Here 
$\cE^{(-1)}$ is an element of sl$(2)$ algebra of $-1$ grade.

Remarkably, the compatibility of the $t_{-1}$ flow with positive $t_n, \, n
\ge 1$ flows does not require
that the matrix $\cE^{(-1)}$ commutes with $E=\l \s_3$, as pointed out in
\ct{Aratyn:2003ym} and \ct{Aratyn:2001pz}.
It turns out that all possible cases are parametrized by a parameter
$\mu$ and fall into two main classes depending on whether $\mu$ takes 
non-zero or zero value.
The corresponding generic choices of $\cE^{(-1)}$ are:
\be
\cE^{(-1)}= \begin{cases} \mu \, \s_3 / {4 \l} 
& \text{for} \;\; \mu \ne 0 \\
\s_{+}/ \l  & \text{for} \;\; \mu  = 0 \, .
\end{cases}
\lab{eminus}
\ee
Note that the value of determinant of $\cE^{(-1)}$ is equal to 
$-\mu^2/16 \l^2 $ and $0$, respectively. There exist other choices of
$\cE^{(-1)}$ for these values of the determinant but they only lead to the 
gauge equivalent copies of hierarchies derived from the choice 
\rf{eminus}.

\section{\sf A class of two-component Schr\"{o}dinger spectral problems}
\label{section:schroedinger}

Consider a linear system 
\br
\psi_{xx}&=&
\left(\frac{\mu^2}{4} - m\,\lambda+\rho^2\,\lambda^2\right)\psi,
\lab{ch-eqx}\\
\psi_t&=&-(\frac{1}{2\l}+u)\,\psi_x+\h u_x \psi \, ,\lab{lch-eqt}
\er
for some arbitrary constant $\mu$ (see \ct{LiuZhang,CH2} for $\mu=1$
and \ct{pavlov} for $\mu=0$).
Compatibility condition for the above system, yields 
three independent equations 
\br
\rho_t  &=& - \( u \rho \)_x \, ,   \lab{cont-eq}\\
m_t &=&- 2 m u_x -m_x u +\rho \rho_x \, ,\lab{mt-eq}\\
m_x &=& \mu^2 u_x - u_{xxx}\, , \lab{mx-def}
\er
corresponding to coefficients of $\l^2$, $\l$ and $\l^0$ in the expansion of 
$\psi_{xxt}-\psi_{txx}=0$. 
Equation \rf{mx-def} can be integrated to yield:
\be
m = \mu^2 u - u_{xx} + \h \kappa \, . \lab{m-def}
\ee
Here, $\kappa$ is an integration constant. For $\mu \ne 0$ that integration 
constant can be removed by transforming the system by Galilean transformation:
\[ x^\pr = x + v t,\;\; t^{\pr} =t , \quad \pder{}{{x^{\pr}}} =\pder{}{x}, \;\;\;
\pder{}{t}= v \pder{}{{x^{\pr}}} + \pder{}{{t^{\pr}}}\, .
\]
In the primed system equation \rf{mt-eq} becomes:
\[ v m_{x^{\pr}}+ m_{t^\pr} = -2 m u_{x^\pr} - m_{x^\pr} u + \rho
\rho_{x^\pr}= -2 \(\mu^2 u - u_{x^\pr x^\pr} + \h \kappa\) u_{x^\pr} - m_{x^\pr} u + \rho
\rho_{x^\pr}
\]
Next, performing a shift $u \to u -v$ and  
choosing velocity $v$ such that  $v=\kappa/2\mu^2$ eliminates the linear terms
in $u_{x}$ and $u_{xxx}$ from the above equation. Clearly, the above
argument works only for $\mu \ne 0$ and from now on we put the integration
constant $\kappa$ to zero as long as $\mu \ne 0$.

Note, that the positive constant $ \mu^2 $, that is different from one 
can be absorbed by appropriately redefining  fields and derivatives. 
Defining ${\wti \l}= \l /\mu^2$, ${\wti \rho}=\rho \mu$, ${\wti u} = u \mu^2$
and new variables ${\wti x}$ and ${\wti t}$ such that 
$\pa_x= \mu \pa_{{\wti x}}$, $\pa_t= (1/\mu) \pa_{{\wti t}}$
allows us to rewrite a linear system \rf{ch-eqx}-\rf{lch-eqt} as
\br
\psi_{{\ti x} {\ti x}}&=&
\left(\frac{1}{4} - m\,{\wti \l} +{\wti \rho}^2\,{\wti \l}^2\right)\psi,
\lab{ch-eqxt}\\
\psi_{\ti t}&=&-(\frac{1}{2{\wti \l}}+{\ti u})\,\psi_{\ti x}+
\h {\wti u}_{\ti x} \psi.\lab{lch-eqtt}
\er
with 
$m= \mu^2 u - u_{xx} + \kappa/2
= {\wti u} - {\wti u}_{{\ti x} {\ti x}}+ \kappa/2$.
Thus, for $ \mu^2 \ne 1$ the spectral system has been transformed to the canonical
system with $ \mu^2 =1$. 

In case of a negative $\mu^2$ (imaginary $\mu$) we make the changes as above 
but with $| \mu^2|$
instead of $\mu^2$ and arrive at
\be
\psi_{{\ti x} {\ti x}}=
\left(-\frac{1}{4} - m\,{\wti \l} +{\wti \rho}^2\,{\wti \l}^2\right)\psi\,.
\lab{ch-eqxtt}
\ee
Thus, only three cases of $\mu^2=-1, 0,1$
need to be considered separately, as concerns equations of motion. 

The case of $\mu^2=1$ corresponds to the 2-component Camassa-Holm model with
$m=u- u_{xx} + \kappa/2$, introduced in \ct{LiuZhang,CH2}, while the case of $\mu^2=-1$ corresponds to 
equation \rf{mt-eq} with $m=-u- u_{xx} + \kappa/2$, which for $\rho=0$
is well-known to possess the compacton solutions \ct{compacton}.

For $\mu=0$, we obtain from \rf{m-def} $m=-u_{xx}+\kappa/2$.
Inserting this value of $m$ into equation \rf{mt-eq} yields
\be
u_{xxt}= -2u_x u_{xx} -u u_{xxx}+ \kappa u_x - \rho \rho_x \, .
\lab{hzeq}
\ee
For $\rho=0$ this is the Dym type equation \rf{dym}.

After one integration (and ignoring the integration constant) we obtain
from \rf{hzeq}
\be
\begin{split}
0&=u_{xt} +u u_{xx}-\kappa u  +\h u_x^2 + \h \rho^2 \\
&=\(u_t +u u_x \)_x - \kappa u +\h \( -u^2_x+ \rho^2\) \, .
\end{split}
\lab{czerou}
\ee
In terms of a new function
\be
v= \h \( -u^2_x+ \rho^2\)
\lab{def-v}
\ee
we can cast equation \rf{czerou} in the following form
\be
\(u_t +u u_x\)_{x} -\kappa u  +v =0 \, .
\lab{gas-a}
\ee
In addition, it follows from equations \rf{czerou} and \rf{cont-eq}
that $v$ defined by relation \rf{def-v} also satisfies 
\be
v_t + \( u (v+u \kappa/2) \)_x = 0 \, ,
\lab{v-cont}
\ee
which becomes a continuity equation in the $\kappa=0$ limit.
The linear system corresponding to equations \rf{gas-a} and \rf{v-cont} 
takes a form
\be
\begin{split}
\psi_{xx}&=
\left( (u_{xx}-\kappa/2) \,\lambda+\(2 v +u_x^2\)\,\lambda^2\right)\psi,\\
\psi_t&=-(\frac{1}{2\l}+u)\,\psi_x+\h u_x \psi\, .
\end{split}
\lab{nl-gas}
\ee
\section{\sf Transformation to the first order spectral problem.
Algebraic Connection to the AKNS model}
\label{section:gauge}

Now, for an arbitrary $\mu$ we perform a reciprocal transformation $(x,t)\mapsto (y,s)$ 
defined by relations
\begin{equation}\lab{reciprocal}
\rd y=\rho\,\rd x-\rho\,u\,\rd t,\quad \rd s=\rd t,
\end{equation}
and 
\begin{equation}\lab{reciporcal-a}
\pder{}{x} =\rho\,\pder{}{y},\quad 
\pder{}{t}=\pder{}{ s}-\rho\,u\,\pder{}{ y}.
\end{equation}
The commutativity of derivatives with respect to $s$ and $y$ variables
is ensured by the continuity equation \rf{cont-eq}.
Applying the reciprocal transformation and then redefining  
$\psi$ by $\varphi=\sqrt{\rho}\,\psi$ as in \ct{CH2} 
leads from the spectral problem 
\rf{ch-eqx}-\rf{lch-eqt} to : 
\br
\varphi_{yy}&=&\left(\l^2-P\,\l-Q \right)\varphi,\lab{zh-specy}\\
\varphi_s&=&-
\frac{\rho}{2\l}\varphi_y+\frac{\rho_y}{4\l}\varphi,\lab{zh-specs}
\er
where
\be
\begin{split}
P&=\frac{m}{\rho^2}\\
Q&=-\frac{\mu^2}{4\rho^2}-\frac{\rho_{yy}}{2\rho}+\frac{\rho_y^2}{4\rho^2} \, .
\end{split}
\lab{pandq}
\ee
Our main point in this section is that we can rewrite the second-order
spectral problem \rf{zh-specy}-\rf{zh-specs} as a first-order linear problem: 
\br
\begin{bmatrix}
\vp \\ \eta \end{bmatrix}_y &=& 
{A} \, \begin{bmatrix} \vp \\ \eta \end{bmatrix} \lab{lspecy}\\
\begin{bmatrix}
\vp \\ \eta \end{bmatrix}_s &=& 
{D} \, 
\begin{bmatrix} \vp \\ \eta \end{bmatrix} \, ,\lab{lspecs}
\er
involving ${\rm sl} (2)$ matrices:
\be
\begin{split}
{ A} &= 
\begin{bmatrix}
g & \l \\ \l -P & -g \end{bmatrix} = \l \s_1+g \s_3 -P \s_{-},\\
{ D} &=
 \frac{1}{\l}  { D}_0 - \h \rho \s_1 , \quad \;\;\;
{ D_0}= - \frac{\mu}{4} \s_3 + \frac{1}{4}\( P \rho - 2 g_s \) \s_{-} \, .
\end{split}
\lab{ad-mat}
\ee
Note that determinant 
of ${D_0}$ is equal to ${\rm det }\, {{D_0}} =-\mu^2/16$ and, therefore,
the matrix $D_0$ becomes singular for $\mu=0$.
Eliminating $\eta$ from the linear spectral problem \rf{lspecy}-\rf{lspecs}
reproduces equations \rf{zh-specy}-\rf{zh-specs}  for $\vp$
providing that function $g(y,s)$ appearing in \rf{ad-mat}
satisfies the Riccati equation 
\[ 
Q= - g^2 -g_y
\]
for $Q$ given in equation \rf{pandq}.
Remarkably, the solution to the above Riccati equation takes a local form of
\be
g (y,s)= \frac{\mu}{2 \rho} + \frac{\rho_y}{2 \rho} \, .
\lab{g-def}
\ee
The zero-curvature equation 
\be
{ A}_s -{ D}_y+\left[ { A}\, ,\, {D} \right] = 0 \, ,
\lab{z-curv}
\ee
can easily be derived from the linear spectral problem \rf{lspecy}-\rf{lspecs}.
It is equivalent to equations:
\be
P_s= \rho_y, \quad \; \; Q_s + \h P_y \rho + P \rho_y=0 \, .
\lab{eqs-zc}
\ee
These equations were found in \ct{CH2} directly from compatibility of equations
\rf{zh-specy}-\rf{zh-specs}.
It follows from the first of the above equations that there exists a 
function $f(y,s)$ such that $ P = f_y$ and $\rho=f_s$.

By plugging $\rho=f_s$ and $P=f_y$ into
the second relation in \rf{eqs-zc} one obtains as in \ct{CH2}:
\be \mu^2 \frac{f_{ss}}{2 f^3_{s}}+
f_{sy}f_y   +\h f_s f_{yy} - \frac{f_{ssyy}}{2f_s}
+\frac{f_{ssy} f_{sy}}{2 f_s^2 } +\frac{f_{ss} f_{syy}}{2 f_s^2 }
-\frac{f_{ss} f_{sy}^2}{2 f_s^3 }=0\,.
\lab{bilcn0}
\ee
This appears to be the only condition, which the function $f$ has to satisfy
in order to be a solution of the model.

For $\mu=0$ equation \rf{bilcn0} simplifies to 
\be
\(f_s^2 f_y - f_{ssy} +\frac{f_{ss}f_{sy}}{f_s} \)_y=0 \, .
\lab{bilc0}
\ee
Integrating the above equation once and setting the integration 
constant to $\kappa/2$ (see explanation below)
yields 
\be
 \frac{f_{ssy}}{f^2_s} -\frac{f_{ss}f_{sy}}{f_s^3} + 
 \frac{\h \kappa}{f_s^2}= f_y
 ,
 \lab{kapeq}
 \ee
 or
\[
 \(\ln f_s\)_{sy} + \h \kappa/f_s = f_s f_y\, .
\]
Indeed, multiplying both sides of equation \rf{kapeq} 
by $f_s^2$ and taking a derivative with respect to $y$ yields
\rf{bilc0}.
It remains to be shown that the choice of $\kappa/2$ as 
the integration constant in equation \rf{kapeq} was consistent with
equations of motion.
To do this we start by recalling that  $P=m /\rho^2$ with 
\[
m =-u_{xx}+\h \kappa= -\rho (\rho u_y)_y +\h \kappa 
\]
in the $\mu=0$ case.
The continuity equation \rf{cont-eq} reads in the hodographic variables
$\rho_s=-\rho^2 u_y$. Accordingly, substituting $ u_y= (1/\rho)_s $ into $P$ we get
\[
P= \frac{(\rho_s/\rho)_y}{\rho} +\frac{\h \kappa}{\rho^2}=f_y \, ,
\]
which is precisely equation \rf{kapeq}.

Let us turn our attention back to the zero-curvature equation \rf{z-curv}.
This equation is invariant under the 
${\rm sl} (2)$ gauge transformation: 
\[
{A} \longrightarrow U {A} U^{-1} + U_y U^{-1}, \quad
{D} \longrightarrow U {D} U^{-1} + U_s U^{-1}\, .
\]
This invariance will be used in what follows to cast 
the linear spectral problem \rf{lspecy}-\rf{lspecs}
in the standard form of the first positive and first negative flow
equations of the ${\rm sl} (2)$ AKNS hierarchy.

As a first step we gauge away the term $ - \h \rho \s_1$ of order $\l^0$
in the expression for ${D}$ in equation \rf{ad-mat} by choosing
\[
U = \exp \biggl( \h f (y,s) \, \s_1 \biggr) =
\cosh \frac{f}{2} +\s_1 \sinh \frac{f}{2} 
\]
so that $U_s U^{-1} -\h \rho \s_1=0$, due to $f_s=\rho$. 
Consequently,
\[
\begin{split}
{{A}} &\to U {{A}} U^{-1} + U_y U^{-1} =
U \( \l \s_1+g \s_3 -P \s_{-} \) U^{-1} +\h P \s_1\\
&=
\l \s_1 +  \s_3 \( g \cosh f - \h P \sinh f \)
-\iu \s_2  \( g \sinh f - \h P \cosh f \) \\
{{D}} &\to \frac{1}{\l} U {{D_0}} U^{-1}  =
\frac{1}{4 \l} \left[ \(- \s_1 \cosh f  - \iu
 \s_2 \sinh f\) \mu \right. \\
& + \left. \(P \rho - 2 g_s \) \( \s_3 +\s_1  \cosh f  + 
\iu \s_2 \sinh f  \) \right]
\end{split}
\]
Note that the gauge transformed of the matrix $D$ is now proportional to 
$1/\l$.

Next, we define the constant matrix 
$\Omega= \frac{1}{\sqrt{2}} \( \s_1 +\s_3\)$, that by a similarity
transformation maps $\s_1$ to $\s_3$, $\Omega \s_1 \Omega^{-1}= -\s_3$. 
Note also that
$\Omega \s_2 \Omega^{-1}= -\s_2$.
The combined gauge transformations first by $U$ and then by 
$\Omega$ produce the final result
\be
\begin{split}
{{A}} &\to E+A_0=\Omega \left[U {{A}} U^{-1} + U_y U^{-1}\right] \Omega^{-1}
\\
&= \l \s_3 + \s_1 \( g \cosh f - \h P \sinh f \) 
+\iu \s_2 \( g \sinh f - \h P \cosh f \)  \\
{{D}} & \to D^{(-1)}=\frac{1}{\l} \Omega U {{D_0}} U^{-1}\Omega^{-1}   =
\frac{1}{4 \l} \biggl[ \( P \rho - 2 g_s \)  \s_3
\biggr.\\&\biggl.+ \s_1 \biggl( \( P \rho -2 g_s \) \sinh f - \mu \cosh f \biggr) 
 +   {\iu}\s_2 \biggl( \(  P \rho -2g_s \) \cosh f - \mu \sinh f
\biggr)  \biggr]
\end{split}
\lab{gt-ad}
\ee
In the above, we re-introduced ${ E}= \l \s_3$ and 
${ A_0}=r \s_{-}+q \s_{+}$. Comparing with the right hand side of equation
\rf{gt-ad} we find that 
\be 
q = e^f \( g - \frac{P}{2} \),\;\; \qquad
r = e^{-f} \( g + \frac{P}{2} \) \, .
\lab{qr-ch}
\ee
Furthermore, defining matrix entries of $D^{(-1)}$ as 
\be
D^{(-1)} = \frac{1}{\l} \begin{bmatrix} \a & \b \\ \g& - \a 
\end{bmatrix}\, , \lab{dinv}
\ee
we find from \rf{gt-ad}, that $\a, \b $ and $\g$ are given by
\be \begin{split}
\a &= \frac{1}{4} \( P \rho - 2g_s \)\\
\b &= e^f \(  \a -\frac{\mu}{4}\) \\
\g &= -e^{-f} \(  \a +\frac{\mu}{4}\) \,.
\end{split}
\lab{alphabg}
\ee
They satisfy the determinant formula $\a^2+\b\g=-{\mu^2}/{16}$.
The matrix entries of $A_0$ and $D^{(-1)}$  enter the 
following simple identities
\begin{alignat}{2}
2 \a&= \b e^{-f} - \g e^{f},&\quad -\frac{\mu}{2} &= \b e^{-f} +\g e^f
\lab{ids-dz} \\
P&= r e^f - q e^{-f},&\quad g &= \h \( r e^f + q e^{-f} \)\, .
\lab{ids-dzaz}
\end{alignat}
It follows that the linear spectral problem \rf{lspecy}-\rf{lspecs} has been
transformed by the above gauge transformation to:
\br
\Psi_y &=&   \( { E} +{ A_0} \) \Psi=\begin{bmatrix} \l & 0 \\ 0& - \l \end{bmatrix} \Psi
+ \begin{bmatrix} 0 & q \\ r& 0 \end{bmatrix}  \Psi \lab{akns-p}\\
\Psi_s &=& D^{(-1)} \Psi= \frac{1}{\l} \begin{bmatrix} \a & \b \\ \g& - \a 
\end{bmatrix}\Psi \lab{akns-n}
\er
for some two-component object $\Psi$. 

We recognize in \rf{akns-p} the spectral problem $L \(\Psi\)=0$ with
the AKNS Lax operator given by equation \rf{lopsl2}. It also follows easily 
that equation \rf{t-1} is the compatibility equation 
of the spectral equations \rf{akns-p}-\rf{akns-n}. 
The compatibility equation \rf{t-1} yields 
\be 
q_s = -2 \beta , \quad r_s = 2 \g  \, .\lab{rsqs}
\ee
when projected on zero grade, and
\be \begin{split}
\a_y &= \h (r q)_s = q \g - r \b \\
\b_y&=-2 \a q\\
\g_y&=2\a r\, ,
\end{split}
\lab{abgy}
\ee
when projected on $-1$ grade. Equations \rf{abgy} can also be directly derived
from equations of motion \rf{eqs-zc}.
\subsection{Examples and Solutions}
\label{subsection:examples}
Let us recall that a
general solution of the compatibility equation \rf{t-1} is given 
by expressions from equation \rf{lrakns}.
It is convenient to 
parametrize the ${\rm SL} (2)$ group element $B$ appearing in expressions
in \rf{lrakns} by the ${\rm sl} (2) $ algebra elements
through the Gauss decomposition:
\be
B = e^{ \chi \s_{-}}\, e^{R \s_3} \,e^{ \p \s_{+}}\, .
\lab{B-def}
\ee
\subsubsection{The case of $\mu \ne 0$}
As an example, we first consider $\mu^2=4$ with $\cE^{(-1)}=\s_3/2\l$
according to equation \rf{eminus}.
As in \ct{Aratyn:2000wr} in order to match the number of independent modes in the matrix $A_0$ we impose
two ``diagonal'' constraints $\Tr \(B_y B^{-1} \s_3 \) = 0$ and
$\Tr \(B^{-1} B_s  \s_3 \)=0$, which effectively eliminate
$R$ in terms of $\p$ and $\chi$.

{}From $B_y B^{-1} =  q \s_{+} +r \s_{-}$ we obtain the 
following representation for $q$ and $r$:
\be
q= -\frac{h_y}{ \Delta} e^R \quad; \quad 
r = - {\bar h}_y \, e^{-R}
\lab{qr-dict}
\ee
where
\be
h = \p \, e^R \quad ; \qquad {\bar h} = \chi \, e^R \qquad ;
\quad \Delta = 1+ h\,{\bar h}
\lab{tipc-def}
\ee
with a non-local field $R$ being determined in terms $h$ and ${\bar h}$ 
from the ``diagonal'' constraints:
\br
\Tr \(B_y B^{-1} \s_3 \) &=& 0 \; \to \; R_y = 
\frac{ {\bar h} h_y}{ \Delta}\, , \lab{par}\\
\Tr \(B^{-1} B_s  \s_3 \) &=& 0 \; \to \; R_s 
= \frac{ h {\bar h}_s }{ \Delta} \, . \lab{bpar}
\er
The zero curvature equations are in this parameterization 
\br
q_s &=&  \(- \frac{h_y}{ \Delta} e^R \)_s
= - 2 h e^R \, ,\lab{LRa}\\
r_s &=&  \(-{\bar h}_y \, e^{-R}\)_s=
- 2 {\bar h} \Delta e^{-R} \, . \lab{LRb}
\er
The two-parameter solution  to the above equations can be deduced 
from a method combining dressing and vertex techniques as described 
e.g. in \ct{Aratyn:1998ef}. The explicit expression is found to be
given by:
\be
\begin{split}
h&= \frac{b \exp \({s /\gamma_2-y\gamma_2}\)}{1+ \Gamma \exp \( s
\( \frac{1}{\gamma_2}-\frac{1}{\gamma_1} \)- y \(\gamma_2-\gamma_1\)\)}\\
{\bar h}&= \frac{a \exp \({-s /\gamma_1+y\gamma_1}\)}{1+ \Gamma \exp \( s
\( \frac{1}{\gamma_2}-\frac{1}{\gamma_1} \)- y \(\gamma_2-\gamma_1\)\)}\\
e^R&= \frac{1+ \frac{\gamma_1}{\gamma_2} \Gamma \exp \( s
\( \frac{1}{\gamma_2}-\frac{1}{\gamma_1} \)- y \(\gamma_2-\gamma_1\)\)}{1+ \Gamma \exp \( s
\( \frac{1}{\gamma_2}-\frac{1}{\gamma_1} \)- y \(\gamma_2-\gamma_1\)\)}\, ,
\end{split}
\lab{solitonhhR}
\ee
where
\[
\Gamma= \frac{ ab \gamma_1\gamma_2}{(\gamma_1-\gamma_2)^2}
\]
is given in terms of four arbitrary constants $a,b,\gamma_1,\gamma_2$.
Higher multi-soliton solutions can be obtained using the same
straightforward procedure.

Comparing with equations \rf{rsqs} we find that
\be  h e^R = e^f \a_{-} , \quad {\bar h} \Delta e^{-R}= e^{-f} \a_{+}
\lab{hhbara}
\ee
where we introduced the notation
\[ \a_{\pm} = 
\a \pm \h \, .
\]
By multiplying the above two relations we find that 
\[ h {\bar h } \D = \( \D-1\) \D = \a_{+} \a_{-} 
= \a^2-\frac{1}{4}
\,.
\]
Solutions to this quadratic equation are
\be
 \Delta = \pm \a_{\pm} , \quad h {\bar h }  =  \D-1 = \pm \a_{\mp} \, .
\lab{solsalpha}
\ee
Adding two relations in \rf{hhbara} we get
\[
2 \a = \a_{+}+\a_{-}= h x+ \frac{1}{x} {\bar h } \D,
\quad x= e^{R-f} \, .
\]
Solving this quadratic equation yields:
\be
f_{\pm} = R +\ln h - \ln \a_{\pm}\, .
\lab{fpm}
\ee
Equation \rf{solsalpha} contains two solutions.
The first one, namely, $\a_{+}=\Delta$, $\a_{-}=\Delta-1$,
when inserted into equation \rf{fpm} yields
\[ f_{+}= R +\ln h - \ln \D, \quad f_{-}= R +\ln h - \ln (\D-1)
=R- \ln {\bar h} \, .
\]
while the second solution,  $\a_{+}=-(\Delta-1)$, $\a_{-}=-\Delta$,
leads to
\[ f_{+}= R +\ln h - \ln (-\D+1), \quad f_{-}= R +\ln h - \ln (-\D)
\, .
\]
Thus, the B\"{a}cklund transformation
\[ f_+ = f_- + \eps \ln \( \frac{\Delta}{\Delta-1} \),\quad  \eps=\pm 1 
\]
relates the two values $f_+$ and $f_-$. 
In the reduced case of sinh-Gordon equation with $h=\bar h$ we find that 
$\Delta=1+h^2$ and
\[
R = \h \ln \( 1+h^2\) = \h \ln \Delta, \quad \ln (\D-1)= 2 \ln h \,.
\]
It follows that all the above values of $f_+, f_-$ can be
summarized as:
\[ f_{\eps} = \frac{\eps}{2} \ln \frac{1+h^2}{h^2}, \quad \eps=\pm 1\, .
\]
\subsubsection{The case of $\mu = 0$}
In the case of $\mu=0$, the matrix $D^{(-1)}$ from equation \rf{dinv}
takes a simple form
\be
D^{(-1)} = \frac{\a}{\l} \begin{bmatrix} 1 & e^f \\ -e^{-f} & - 1 
\end{bmatrix}, \lab{dinvcz}
\ee
which according to definition \rf{eminus} is reproduced by
\[
\frac{1}{\l} B \s_+ B^{-1}= \frac{1}{\l} \begin{bmatrix} - \chi & 1 \\  -\chi^2 & \chi 
\end{bmatrix} e^{2 R}
\]
for
\be \begin{split} 
\chi&= - e^{-f} \\
e^{2 R} & = \a e^f
\end{split}
\lab{chirpara}
\ee
or $R= \( f + \ln \a \)/2$.

{}From equation \rf{abgy} we find for $\mu=0$ that
$\a_y= -2 \a g$. Therefore
\be
g = - \frac{\a_y}{2 \a} = - \h (\ln \a)_y
\lab{gczero}
\ee
and by comparing with definition \rf{g-def} we conclude that
\be
\( {\rho}{\a}\)_y=0 \, .
\lab{rhoa}
\ee

Next, we calculate 
\be
B_y \, B^{-1} = \( R_y- \chi \p_y e^{2 R} \) \s_3 
+ \p_y e^{2R} \s_{+} + 
\( \chi_y +2 \chi R_y - \chi^2 \p_y e^{2R} \) \s_{-} \, .
\lab{bybinv}
\ee
Imposing condition $\Tr \(B_y B^{-1} \s_3 \) = 0$ implies
$R_y- \chi \p_y e^{2 R} =0$ or
\[
\p_y= -R_y/\a= - \frac{1}{2 \a} \( f_y + \frac{\a_y}{\a}\)
= \frac{1}{\a} \( g - P/2\) \, .
\]
What remains of expression \rf{bybinv} is now given by   
\[
B_y \, B^{-1} = 
 \p_y e^{2R} \s_{+} + 
\( \chi_y + \chi R_y \) \s_{-}=
( R_y/\chi) \s_{+} + 
\( \chi_y + \chi R_y \) \s_{-}
\, .
\]
Recalling relations \rf{chirpara} and \rf{gczero} we obtain
the desired results
\[ 
q= \p_y e^{2R} = \( g - \h P\) e^f , \quad \; 
r=\chi_y + \chi R_y= \( g + \h P\) e^{-f}\, ,
\]
that reproduce expressions \rf{qr-ch}. 

The compatibility equation
\[
\( B_y \, B^{-1}\)_s = \lb B \s_{+} B^{-1}, \s_3 \rb
=\begin{bmatrix} 0 & -2 \\
-2 \chi^2 & 0 \end{bmatrix} e^{2R}
\]
yields the following equations of motion
\br
\(\frac{R_y}{\chi}\)_s &=& -2 e^{2R} \lab{ryxs1}\\
\chi_{ys}+2 \chi_sR_y &=& 0\;\; 
\to \;\; \(\chi_s e^{2R}\)_y = 0 \, .\lab{ryxs2}
\er
Equation \rf{ryxs2} implies that
\be
\chi_s = c_3 (s) \, e^{-2 R} \, ,
\lab{chisc3}
\ee
where $c_3(s)$ is a an arbitrary function of $s$ only.

{}From equations \rf{chirpara} we find that 
\[
\chi_s = f_s e^{-f}= \rho e^{-f}= c_3 (s) \a^{-1} e^{-f}
\]
and therefore $ \rho = c_3(s) / \a$ in agreement with relation 
\rf{rhoa}. It follows that $c_3 (s)$ has to be different from zero for
consistency of the model with $\rho \ne 0$.

Integrating relations \rf{chisc3} and \rf{ryxs1} leads to 
\br
\chi &= & \int^s c_3 (s) \, e^{-2 R} \, \rd s + c_2 (y) \lab{chiint}\\
\frac{R_y}{\chi} &= &-2  \int^s  e^{2 R} \, \rd s + c_1 (y)\,,  \lab{Rychiint}
\er
where $c_1 , c_2$ depend at most on $y$. Combining these two equations
and setting $c_3 (s)$ to be a constant $c_3$ we get
the deformed sinh-Gordon equation for $R$ \ct{zenchuk}:
\be
R_y= -2 c_3  \int^s  e^{-2 R} \, \rd s \, \int^s  e^{2 R} \, \rd s
-2 c_2  \int^s  e^{2 R} \, \rd s + c_1 c_3 \int^s  e^{-2 R} \, \rd s
+ c_1 c_2
\lab{deformsg}
\ee
or
\[
R_{ys}= -2 c_3  \(\int^s  e^{-2 R} \, \rd s \, \int^s  e^{2 R} \, \rd s\)_s
-2 c_2  e^{2 R}+ c_1 c_3   e^{-2 R} \,.
\]
The one soliton solution to the above equation with $c_1=c_2=0$ is given by
(see also \ct{zenchuk})
\be
R(s,y)= \frac{s}{2} + 2 c_3 y +\ln \( 
\frac{k_0 e^{k_1 s +k_2 y}+k_1+1}{k_0 e^{k_1 s +k_2 y}-k_1+1} \)\, ,
\lab{Rsoliton}
\ee
where 
\[ k_2 = 8 c_3 \frac{k_1}{k_1^2-1}
\]
and $k_0$, $k_1$ are real constants. The corresponding expression for $\chi$
is
\be
\chi(s,y) = - e^{-f} = -e^{-s-4 c_3 y}\,
\frac{(k_1-1)^2/(k_1+1) + k_0 e^{k_1 s +k_2 y}}{k_0 e^{k_1 s +k_2 y}+k_1+1}
c_3 \, .
\lab{chif}
\ee
The above function $\chi$ together with $R$ from \rf{Rsoliton}
solve equations \rf{ryxs1}-\rf{ryxs2}.
The function $f$ defined by equation \rf{chif} 
provides a one-soliton solution to equation \rf{bilc0}.
It satisfies
\[
f_s = \frac{c_3}{\a}
\]
with
\be
\a = \frac{\((k_1-1)^2/(k_1+1) + k_0 e^{k_1 s +k_2 y}\)\(
k_0 e^{k_1 s +k_2 y}+k_1+1\)}{\(
k_0 e^{k_1 s +k_2 y}-k_1+1\)^2}  c_3\, .
\lab{alpha-sol}
\ee
Eliminating $\a$ from equation \rf{ryxs1} using \rf{chirpara}
we get
\[
-\(\ln f_s\)_{ys} + f_s f_y= 4 \a= \frac{4 c_3}{f_s} \, .
\]
Therefore, comparing with \rf{kapeq}, we see that $\kappa=8 c_3$ and $f$
given in \rf{chif} satisfies \rf{kapeq} and therefore also \rf{bilc0}.
One-soliton solution $R(s,y)$ given by expression \rf{Rsoliton}
satisfies therefore:
\[
R_y= -\frac{\kappa}{4}  \int^s  e^{-2 R} \, \rd s \, \int^s  e^{2 R} \, \rd
s \, .
\]

\section{\sf Bihamiltonian structure}
\label{section:bihamiltonian}
\subsection{Bihamiltonian structure of the 2-component Camassa-Holm model}
As in \ct{LiuZhang}, we consider the following bihamiltonian structure:
\be
\begin{split}
\{w_1(x), w_1(x')\}_1&=\{w_2(x), w_2(x')\}_1=0,\\
\{w_1(x), w_2(x')\}_1&=\delta'(x-x')-\frac{1}{\mu} \delta''(x-x').\\
\{w_1(x), w_1(x')\}_2&=2 \delta'(x-x'),\\
\{w_1(x), w_2(x'))\}_2&=w_1(x) \delta'(x-x')+w_1'(x) \delta(x-x'),\\
\{w_2(x), w_2(x')\}_2&= w_2(x)\d^{\pr} (x-x') +\pa_x \(w_2(x) \delta(x-x')\)\, ,
\end{split}
\lab{bihamstr}
\ee
where $1/\mu$ now plays a role of the deformation parameter.
There exists an hierarchy of Hamiltonians related through 
Lenard relations \ct{LiuZhang}
\begin{equation}
\{w_i(x), H_{-j}\}_2=j \{w_i(x), H_{-j-1}\}_1,\quad\;\;\; j=1,2,3,{\ldots} \, .
\lab{lenard}
\end{equation}
The flows of the bihamiltonian hierarchy are generated
by the Hamiltonians $H_{-j}, j< 0$ via:
\begin{equation}\lab{hamflows}
\pder{w_i}{t_{-j+2}}=\{w_i(x), H_{-j}\}_1,\quad j=3,{\ldots} ,\; i=1,2 \, .
\end{equation}
The lower Hamiltonians $H_{-j}$ for $j>1$ can be obtained recursively 
from the Casimir $H_{-1}=\int w_2(x) dx$ of the first Poisson bracket
applying the Lenard relations \rf{lenard}.
Following \ct{LiuZhang,falqui}, we introduce objects 
$\varphi_1, \varphi_2$ defined by $ w_1=\varphi_1-  \varphi_{1,x}/\mu,
\;w_2=\varphi_2+ \varphi_{2,x}/\mu$.
Then 
\[ \{\vp_1(x), w_2(x')\}_1=\delta'(x-x'), \quad 
\{w_1(x), \vp_2(x')\}_1=\delta'(x-x')
\]
and the Lenard relations yield
\br
H_{-2} &=& \int \lb \vp_2 \( \vp_1- \vp_{1, x}/\mu \)\rb \rd x \nonu\\
H_{-3} &=&\h \int\lb \vp_2^2 +\vp_2\vp_1 \( \vp_1- \vp_{1, x}/\mu \)\rb \rd x
\, .\nonu
\er
Plugging the above $H_{-3}$ into equation of motion \rf{hamflows} for $j=3$ 
we obtain: 
\be \begin{split}
(w_1)_{t}&=\(\varphi_2+\h
\varphi_1^2-\frac{1}{2\mu} \varphi_1 \varphi_{1,x}\)_x,  \\
(w_2)_{t}&=\(\varphi_1 \varphi_2+\frac{1}{2\mu} \varphi_1 \varphi_{2,x}\)_x \, ,
\end{split}
\lab{2-ch}
\ee
where $t=t_{-1}$.
Defining $u$ such that $\vp_1=2 u$ and $\rho$ such that 
$w_2=-\rho^2/\mu^2 +w_1^2/4$ or $\rho^2=w_1^2 \mu^2/4 -w_2\mu^2$
we can rewrite the above system of equations after a transformation $t \to -t$
as 
\begin{eqnarray}
u_t - u_{xxt}/\mu^2&=&
\rho \rho_x/\mu^2 - 3 u u_x + 2 u_x u_{xx}/\mu^2 + u u_{xxx}/\mu^2
\lab{epsch1}\\
\rho_t &= &-(u \rho)_x \, ,\lab{epsch2}
\end{eqnarray}
which agrees with the 2-component Camassa-Holm equation.
Multiplying equation \rf{epsch1} by $\mu^2$ and taking $\mu^2 \to 0$
yields 
\be
-u_{xxt}=
\rho \rho_x + 2  u_x u_{xx} + u u_{xxx}
\lab{epsch3}
\ee
corresponding to eq. \rf{hzeq} with $\kappa=0$.

In order to take the $\mu \to 0$ limit of the Poisson structure
\rf{bihamstr} it is convenient to change the
variables from $w_1$, $w_2$ to $m$ and $\rho$ defined as
\be
\begin{split}
m  &= \h \mu^2 \( w_1 (x) + w_{1, x} / \mu\) = \mu^2 u - u_{xx} \\
\rho^2&= \mu^2 \(w_1^2 /4 -w_2\) \, .
\end{split}
\lab{mrhovar}
\ee
In terms of $m$ and $\rho$ the Poisson bracket structure \rf{bihamstr}
turns into:
\be
\begin{split}
\{m (x), m (x')\}_1&=0,\\
\{\rho^2 (x), \rho^2 (x')\}_1&= -\mu^2 \bigl( 2 m(x) \d^{\pr} (x-y)
+m^{\pr} (x) \d (x-y) \bigr),\\
\{\rho^2 (x), m (x')\}_1&=\h \mu^2 \( - \mu^2 \delta'(x-x')+
\delta^{\pr\pr\pr} (x-x')\),\\
\{m (x), m (x')\}_2&=\h \mu^2 \(  \mu^2 \delta'(x-x')-
\delta^{\pr\pr\pr} (x-x')\),\\
\{\rho (x), \rho (x'))\}_2&=-\h \mu^2 \delta^{\pr} (x-x'),\\
\{\rho (x), m (x')\}_2&= 0\, ,
\end{split}
\lab{bihamrhom}
\ee
\subsection{The $\mu \to 0$ limit and the Dym type hierarchy.}
Redefining the brackets as follows
\[ 
\{ \cdot , \cdot \}_j \; \longrightarrow \; \mu^2 \{ \cdot , \cdot \}_j 
\]
and taking $\mu \to 0$ limit in equation \rf{bihamrhom}
we find for the first and second bracket
structure in terms of $u$ and $\rho$ (see also \ct{pavlov,dn}):
\be
\begin{split}
\{u(x), u (x')\}_1&=0,\\
\{ \rho^2 (x), \rho^2 (x')\}_1&= 
 2 u^{\pr \pr}(x') \d^{\pr} (x-x') - u^{\pr \pr\pr} (x')\d (x-x') \\
\{ \rho^2 (x), u (x')\}_1&=- \h \d^{\pr} (x-x'),\\
\{u (x), u (x')\}_2&=- \h \pa^{-1}_x \delta(x-x'),\\
\{\rho (x), \rho (x')\}_2&=- \h \pa_x \delta(x-x'),\\
\{u (x), \rho  (x')\}_2&= 0\, ,
\end{split}
\lab{urho2c0}
\ee
The first bracket in \rf{urho2c0} has the Casimir:
\[
H_{-1}^{(1)}=  \int \lb \rho^2 (x)-u_x^2 (x) \rb \rd x\,.
\]
This Casimir leads via Lenard relation \rf{lenard} to the Hamiltonian:
\[
H_{-2} = 2 \int (\rho^2 -u_x^2)\, u \rd x \, ,
\]
which in turn generates equations of motion of $\mu=0$ case via equations \rf{hamflows}
and \rf{lenard}:
\be
\begin{split}
\pder{u_x}{t} &= \h \{ u_x (x) , H_{-2} \}_2= - \h (u_x)^2 -u u_{xx}
-\h \rho^2 \\
\pder{\rho}{t} &= \h \{ \rho (x) , H_{-2} \}_2= - \( u \rho \)_{x}
\lab{hamflowsc02}
\end{split}
\ee
This Hamiltonian structure can be extended by an 
additional term:
\[
{\bar H}_{-2} = -2 \kappa \int u^2 \rd x \, .
\]
Adding this term to $H_{-2}$ will lead via 
relations \rf{hamflowsc02} to correct equations
of motion \rf{gas-a}-\rf{v-cont}.

\subsection{Hamiltonians of positive order}
There exists another class of conserved charges, different from the chain of
Hamiltonians $H_{-j},\, j=1,2,{\ldots}  $ of negative order discussed above.
These are the Hamiltonians of positive order originating from the
Casimir 
\be
H_{-1}^{(2)}= 2 \int \rho (x) \rd x
\lab{casim2}
\ee
of the second Poisson bracket \rf{bihamrhom}.
We now employ Lenard relations \rf{lenard} 
to construct higher order Hamiltonians.
The first recurrence step:
\[ \{ \cdot , \int \rho (x) \rd x \}_1 = \{ \cdot , H_0 \}_2 \, ,
\]
where ``$\cdot$'' stands for phase space variables 
$m(x)$ and $\rho (x)$, leads to a new Hamiltonian:
\be
H_0 =  - \int \frac{m(x)}{\rho (x)} \rd x \, .
\lab{hzero}
\ee
in 
agreement with expression found in \ct{falqui}. 
The integrand of $H_0$ can be rewritten as
\[
\frac{m}{\rho }= \rho \frac{m}{\rho^2}= \rho P = \rho f_y
=f_x
\]
and therefore $H_0$ appears to be a surface term that would vanish if $f$ 
would be a local field.

On the next level we find from the Lenard relations: 
\[ \{ \cdot , H_0 \}_1 = \{ \cdot , H_1 \}_2 \, ,
\]
with 
\be
H_1 = \int \big\lb \frac{1}{4 \rho^3} \( \rho_x^2-m^2\)+\frac{\mu^2}{4 \rho} \big\rb
\rd x
\lab{hone}
\ee
This recurrence process can be continued to yield higher order
Hamiltonians. Technical calculations 
involved in obtaining higher order Hamiltonians become 
increasingly tedious.
Remarkably, we can bypass these difficulties by relying on 
the underlying AKNS structure governing higher positive flows.
We recall the Hamiltonian densities 
$\cH_n$ \rf{17a} of the AKNS model generating the positive flows
of the model. Their conservation law with respect to
the negative flow $s$ takes a form
\be
\(\cH_{n}\)_{ s} = X_{y} \, ,
\lab{akns-con}
\ee
where $s$ and $y$ are ``reciprocal'' variables describing time and space
of the AKNS model and $X$ is some local quantity. The above relation
ensures that $(\int \cH_{n}\rd y)_s=0$ (for $X$ local in $u$ and $\rho$)
and thus the integral $\int \cH_{n}\rd y$ is conserved.

In terms of the original $t,x$ variables the conservation
laws \rf{akns-con} read 
\[
\( \pder{}{t} + \rho u \pder{}{y}\) \cH_{n } = \pder{}{y} X
\]
or
\[
 \pder{}{t} \cH_{n } = - \rho u \pder{}{y}   \cH_{n }
 + \pder{}{y} X =  -  u \pder{}{x}   \cH_{n }
+ \frac{1}{\rho} \pder{}{x} X \, ,
\]
where we used that $\pa/\pa y = \rho \pa/\pa x$.
It follows that 
\be
\pder{}{t} \( \rho \cH_n\) = - (u \rho)_x    \cH_{n}
- u \rho \cH_{n \, x} + X_x= \pder{}{x} \( X - u \rho H_{n} \)\, .
\lab{aknsc1}
\ee
Thus, the quantities $\rho \cH_n$ are conserved charges of the 2-component
Camassa-Holm and 2-component Dym type models.

The first two Hamiltonian densities $rq$ and $r q_y$ of the AKNS model,
given by relations \rf{18a} and \rf{20a},
give rise, after use of definitions \rf{qr-ch},\rf{pandq} and \rf{g-def}, 
to the following conserved charges:
\be
\begin{split}
H_1&=\int \rho \cH_1 \rd x = \int \rho r q \rd x = 
\int \big\lb \frac{1}{4 \rho^3} \( \rho_x^2-m^2\)+\frac{\mu^2}{4 \rho} \big\rb
\rd x,\\
H_2&= \h \int \rho \cH_2 \rd x  =\h \int  \rho r q_y \rd x=
\h \int  r q_x \rd x\\&= 
\int \frac{m}{2\rho^2} \bigg\lb \frac{\mu^2}{4 \rho} - \frac{3}{4}
\frac{\rho_x^2}{\rho^3} +\frac{\rho_{xx}}{2 \rho^2} -\frac{m^2}{4 \rho^3}\bigg\rb
\rd x\, ,
\end{split}
\lab{rhohams}
\ee
which we have verified explicitly to be conserved under equations of motion \rf{cont-eq}-\rf{mt-eq}
for $m=\mu^2 u-u_{xx}$ for $\mu \ne 0$ and $m=-u_{xx}+\kappa/2$ for $\mu=0$.

We recognize in $\rho \cH_1$ the Hamiltonian $H_1$
derived in \rf{hone} from the Casimir of the second bracket
via Lenard recursion relations.
Furthermore, we have shown that $H_1= \int \rho \cH_1\rd x$ 
and $H_2=\int \rho \cH_2\rd x/2$  also are interrelated via 
the Lenard relation:
\[
 \{ \cdot , H_1  \}_1= \{ \cdot ,H_2 \}_2.
\]
Therefore, the conclusion is that the AKNS induced Hamiltonians 
$\rho \cH_n$ form the sequence of positive order Hamiltonians
of the 2-component Camassa-Holm and 2-component Dym type hierarchies.
The formula \rf{17a} given in section \ref{section:akns} can be 
used to systematically derive all the Hamiltonians governing positive flows 
of this model.

\section*{\sf Acknowledgments}
We thank A. Das and L.A. Ferreira for helpful discussions.
H.A. acknowledges support from Fapesp and thanks the IFT-UNESP Institute
for its hospitality. JFG and AHZ thank CNPq for a partial support.

\end{document}